\documentclass[usenatbib]{mn2e}

\usepackage{verbatim}
\usepackage{graphicx}
\usepackage{url}
\usepackage{amsfonts}
\usepackage{amsmath,amssymb}

\newcommand{\beqs} { \begin{eqnarray} }
\newcommand{\eeqs} { \end{eqnarray} }
\newcommand{\bsub}{ \begin{subequations} }
\newcommand{\esub}{ \end{subequations} }

\renewcommand{\eqref}[1]{(\ref{#1})}

\newcommand{\ergs}{\mbox{erg\,s$^{-1}$}}

\newcommand{\lgrav}{\mbox{$\cal L_{\rm G}$}}
\newcommand{\ls}{\mbox{$\cal L_*$}}
\newcommand{\rs}{\mbox{$R_{\rm S}$}}

\title{On the stellar luminosity of the universe}
\author[Ralph A.M.J. Wijers]{Ralph A.M.J. Wijers\thanks{E-mail:
rwijers@science.uva.nl} \\
Astronomical Institute Anton Pannekoek, Faculty of Science, University
of Amsterdam and \\
Centre for High-Energy Astrophysics, Kruislaan 403, 1098 SJ Amsterdam, The Netherlands}
\begin{document}

\date{}

\pagerange{\pageref{firstpage}--\pageref{lastpage}} \pubyear{2005}

\maketitle

\label{firstpage}

\begin{abstract} 
It has been noted at times that the rate of energy release in
the most violent explosive events in the Universe, such as
supernovae and gamma-ray bursts, rivals the stellar luminosity of
the observable universe, \ls. The underlying reason turns out to
be that both can be scaled to $c^5/G\equiv\lgrav$, albeit that for
the explosions $L/\lgrav$ follows from first principles, whereas
for \ls\ the scaling involves quantities too complex to derive
from elementary considerations at the present time.
Under fairly general circumstances, \ls\ is dominated
by stars whose age is similar to the Hubble time.
\end{abstract}

\begin{keywords}
cosmology -- stellar evolution -- gamma-rays: bursts -- supernovae
\end{keywords}


\section{Introduction}

When a massive star dies, its core collapses to, or close to,
the Schwarzschild radius, $\rs\equiv 2GM/c^2$. Since of the order
of the rest energy of the mass can thus be released, in a time
close to the dynamical time ($\rs/c$), this can lead to the
release of energy at a rate $Mc^2/(\rs/c)\sim c^5/G\equiv\lgrav
(=3.7\times10^{59}\ergs)$.
This number is independent of the collapsing mass, and might be 
achieved in practice, for example, as the gravity-wave luminosity 
from two merging black holes\footnote{
  As is well known, $\lgrav$ is also the only
  quantity of dimension luminosity that can be constructed out of
  the fundamental constants of gravity and relativity, and thus the
  only natural luminosity in the large-scale world (as that is
  uncharged and not quantum-mechanical).
}.
Since gamma-ray bursts (at least of the long-soft variety) are thought
to result from the core collapse of massive stars, it is not surprising
that their luminosity should be a fair fraction of $\lgrav$ (for
a review on gamma-ray bursts, see \cite{Paradijs2000}). Even the deviation
from $\lgrav$ can be understood quantitatively. For example, the
neutrino luminosity of a supernova is down by a factor 0.1 because only
that fraction of the rest mass is released, and by another factor $10^5$
because the neutrinos are released on the much slower diffusion time rather
than the dynamical time, leading to a luminosity of order $10^{53}\ergs$.

Is it, however,
understandable that this luminosity should roughly rival the stellar
luminosity, $\ls$, of the observable universe, or is that a mere numerical
coincidence? A rough estimate puts $\ls$ at a few times $10^{55}$\ergs, since
the observable universe contains of order $10^{11}$ galaxies, each with on
average $10^{11}$ stars, that emit a few times $10^{33}$\ergs\ per star.
In sect.~\ref{sec:ls} I show that one can actually write $\ls$ in
terms of fundamental quantities, albeit that some microscopic processes
enter via parameters; in sect.~\ref{sec:ts} I then eliminate the
dependence on the stellar nuclear burning time. The main findings
are summarized in sect.~\ref{sec:concl}.

\section{The luminosity of all stars}
\label{sec:ls}

In this work I am after finding characteristic numbers and global scalings,
not precise values. I therefore do not calculate corrections for and
integrations over redshift, but simply count the observable universe as
the Euclidean volume out to the Hubble radius, $r_{\rm H}=ct_{\rm H}$.
I also approximate $t_{\rm H}=H_0^{-1}$, which is strictly valid only
for an empty Universe, but close enough for plausible others.
(For these, and other elementary cosmology relations, see, e.g.,
\cite{Peebles1993}.)

First we express the luminosity of one star
in terms of its mass and lifetime:
\begin{equation}
L_* = \eta \frac{M_*c^2}{t_*}
\end{equation}
Since we are going to average over stellar populations eventually, we are
only interested in the average luminosity and count $t_*$ as its total
lifetime, rather than the instantaneous nuclear burning time of a specific
evolutionary phase. This in turn is well approximated by the main-sequence
lifetime. $\eta$ is the nuclear burning efficiency, which is 0.007 for
the fractional mass loss due to fusion, times another factor 0.5 to account
for the fact that only about half the initial mass of a star will go
through nuclear burning.
Assuming that a fraction $f_*$ of all the mass within the Hubble radius,
$M_{\rm H}$, is in stars, we can write the total stellar luminosity
within the Hubble radius as
\begin{equation}
\ls = \eta f_* M_{\rm H} c^2/t_*.
\end{equation}

In order to proceed from here, it helps to note that the Schwarzschild
radius of all the matter within the Hubble radius is of order 
$\Omega r_{\rm H}$. This can be seen as follows:
\begin{equation}
R_{\rm S,H} = \frac{2GM_{\rm H}}{c^2} = \frac{8\pi G}{3c^2} \rho_{\rm m} (ct_{\rm H})^3,
\end{equation}
where we have made the Euclidean-volume approximation, and $\rho_{\rm m}$
is the total matter density (baryonic plus dark). Now we can write
$\rho_{\rm m} = \Omega_{\rm m}\rho_{\rm c}$, where the critical density, $\rho_{\rm c}=
3H_0^2/8\pi G$, and use $H_0t_{\rm H}=1$ to get
\begin{equation}
R_{\rm S,H} = \Omega_{\rm m} ct_{\rm H} = \Omega_{\rm m} r_{\rm H},
\end{equation}
as stated at the start. (Somewhat crudely, one could say that a flat universe
comes close to living inside its own black hole.)
We can now use this result to rewrite $M_{\rm H}$
in the expression for \ls:
\begin{equation}
M_{\rm H} \equiv \frac{c^2R_{\rm S,H}}{2G} = \frac{c^3\Omega_{\rm m} t_{\rm H}}{2G}.
\end{equation}
Putting everything together, we then get 
\begin{equation}
\ls = \frac{f_*\eta\Omega_{\rm m}}{2}\frac{t_{\rm H}}{t_*}\frac{c^5}{G} =
      \frac{f_*\eta\Omega_{\rm m}}{2}\frac{t_{\rm H}}{t_*}\lgrav.
\end{equation}
Recalling that $\eta\simeq1/300$, $\Omega_{\rm m}\simeq0.3$, and about 10\% of the
baronic mass is in stars\footnote{
   Recall that $f_*$ is defined as the mass in stars divided by the total
   (baryonic plus dark) matter mass.
} (e.g., \cite{Fukugita2004}), 
so $f_*\sim0.01$, this implies $\ls\sim
10^{55}t_{\rm H}/t_*$. This is in reasonable agreement with the crude
observational estimate, provided that the relevant stellar age is close
to the Hubble time.

\section{The age of the dominant stars}
\label{sec:ts}

The next step is to show that indeed usually the relevant mean stellar age
equals the Hubble time. Note again that we average over scales  that are
a significant fraction of the horizon scale, so that we parametrize
the average of all types of star formation (from slow and gradual to massive
starbursts) simply by what is the total mass in stars, and what is the
representative mean initial mass function (IMF) of stars.
Let the formation rate of stars of mass 
$M$ at time $t$ since the Big
Bang be given by
\begin{equation}
S(M,t) = Kt^{-q}M^{-1-x},
\end{equation}
where $x\simeq1.5$ \citep{Salpeter1955,Miller1979},
 and $q$ is positive, but not large, currently 
(i.e., we take $S$ simply to be proportional to the (time-independent)
IMF and to change gradually with time). The 
present day mass function $P(M)$ can then be found by integrating
$S(M,t)$ over time. For small masses, all stars formed since the
beginning are still around, and so we simply integrate the time dependence.
For large masses, only the stars formed during a time $t_*(M)\propto M/L$
in the recent past count. The transition between the two regimes happens
at the mass $\tilde{M}$, for which $t_*(\tilde{M})=t_{\rm H}$, so we have
\begin{eqnarray}
P(M) & \propto &  \makebox[3cm][l]{$t^{1-q} M^{-1-x}$}   M<\tilde{M}\nonumber \\
     & \propto &  \makebox[3cm][l]{$t^{-q} M^{-x}/L(M)$}  M>\tilde{M}.
\end{eqnarray}
To get the contibution to the total luminosity from mass $M$, we
finally multiply by $M.L(M)$:
\begin{eqnarray}
\cal{L}(M) & \propto &  \makebox[3cm][l]{$t^{1-q} M^{-x} L(M)$} M<\tilde{M}\nonumber \\
     & \propto &        \makebox[3cm][l]{$ t^{-q} M^{1-x}$}     M>\tilde{M}.
\end{eqnarray}
Since $L$ scales very steeply with $M$, typically $L\propto M^{3.5}$ on
the lower main sequence, this implies that ${\cal L}\propto M^2$ up to
$\tilde{M}$, and ${\cal L}\propto M^{-0.5}$ thereafter, so indeed the
total luminosity from a gradually-forming population of stars is 
dominated by those for which $M=\tilde{M}$, hence $t_*=t_{\rm H}$. 
Note that this result stems
from the steep negative slope of the initial mass function combined with the 
steep positive slope of the mass-luminosity relation for stars. Specially,
for a significantly flatter IMF than the present-day $x\simeq1.5$, the
most massive stars in the population might dominate the total luminosity,
and then the result is not valid. With that caveat, however, we may conclude
that under conditions of star formation as we now know it, it is valid
to set $t_*=t_{\rm H}$ and thus
\begin{equation}
 \label{eq:lstar}
 \ls \simeq f_*\eta\Omega_{\rm m}\lgrav.
\end{equation}

\section{Conclusions}
\label{sec:concl}

Starting from the nuclear burning rate of a star, and using Euclidean
approximations to get rough numbers for the mass in the observable
universe, I find that the total stellar luminosity in the observable universe
can be expressed as a fraction of the characteristic gravitational luminosity,
$\lgrav=c^5/G$. The constants of proportionality in the final equation,
eq.~\ref{eq:lstar}, hide our ignorance about the details of star formation
in the form of $f_*$, the fraction of mass in the universe that is in
stars. Part of the reason why a simple expression of this type could be 
found is that both the birth rate of stars and the luminosity of a star
are steep functions of stellar mass, which conspire to make the stars
that dominate the total luminosity to be always those that are about the
same age as the universe (or of any other closed system in which star
formation is a slow function of time).

\label{lastpage}
\end{document}